\documentclass[final]{svjour3}
\usepackage{graphicx}
\usepackage{rotating}
\usepackage{multirow}
\usepackage{amssymb}
\usepackage{mathptmx}
\usepackage{siunitx}
\usepackage{hyperref}
\usepackage[square,numbers]{natbib}
\usepackage{xcolor}
\newcolumntype{C}{>{\centering\arraybackslash}p{5em}}
\makeatletter
\journalname{Journal of Low Temperature Physics}

\begin{document}

\newcommand{\hdblarrow}{H\makebox[0.9ex][l]{$\downdownarrows$}-}

\title{Progress Report on the Large Scale Polarization Explorer}

\authorrunning{L. Lamagna et al.}
\titlerunning{Progress Report on the Large Scale Polarization Explorer}

\author{L.~Lamagna$^{1,2,*}$ \and G.~Addamo$^{3}$ \and P.~A.~R.~Ade$^{4}$\and C.~Baccigalupi$^{5}$ \and A.~M.~Baldini$^{6}$ \and P.~M.~Battaglia$^{7}$ \and E.~Battistelli$^{1,2}$ \and A.~Ba\`{u}$^{8}$ \and M.~Bersanelli$^{9,10}$ \and M.~Biasotti$^{11,12}$ \and C.~Boragno$^{11,12}$ \and A.~Boscaleri$^{13}$ \and B.~Caccianiga$^{10}$ \and S.~Caprioli$^{9,10}$ \and F.~Cavaliere$^{9,10}$ \and F.~Cei$^{6,14}$ \and K.~A.~Cleary$^{15}$ \and F.~Columbro$^{1,2}$ \and G.~Coppi$^{16}$ \and A.~Coppolecchia$^{1,2}$ \and D.~Corsini$^{11,12}$ \and F.~Cuttaia$^{17}$ \and G.~D'Alessandro$^{1,2}$ \and P.~de~Bernardis$^{1,2}$ \and G.~De~Gasperis$^{18,19}$ \and M.~De~Petris$^{1,2}$ \and F.~Del~Torto$^{20}$ \and V.~Fafone$^{18,19}$ \and Z.~Farooqui$^{3}$ \and F.~Farsian$^{5}$ \and F.~Fontanelli$^{11,12}$ \and C.~Franceschet$^{9,10}$ \and T.C.~Gaier$^{21}$ \and F.~Gatti$^{11,12}$ \and R.~Genova-Santos$^{22}$ \and M.~Gervasi$^{8,23}$ \and T.~Ghigna$^{24}$ \and M.~Grassi$^{6}$ \and D.~Grosso$^{11,12}$ \and R.~Gualtieri$^{32}$ \and F.~Incardona$^{9,10}$ \and M.~Jones$^{24}$ \and P.~Kangaslahti$^{15}$ \and N.~Krachmalnicoff$^{5}$ \and R.~Mainini$^{8}$ \and D.~Maino$^{9,10}$ \and S.~Mandelli$^{9,10}$ \and M.~Maris$^{25}$ \and S.~Masi$^{1,2}$ \and S.~Matarrese$^{26}$ \and A.~May$^{27}$ \and P.~Mena$^{28}$ \and A.~Mennella$^{9,10}$ \and R.~Molina$^{28}$ \and D.~Molinari$^{29,30}$ \and G.~Morgante$^{17}$ \and F.~Nati$^{8}$ \and P.~Natoli$^{29,30}$ \and L.~Pagano$^{29,30}$ \and A.~Paiella$^{1,2}$ \and F.~Paonessa$^{3}$\and A.~Passerini$^{8}$ \and M.~Perez-de-Taoro$^{22}$ \and O.~A.~Peverini$^{3}$ \and F.~Pezzotta$^{9,10}$ \and F.~Piacentini$^{1,2}$ \and L.~Piccirillo$^{27}$ \and G.~Pisano$^{4}$ \and L.~Polastri$^{29,30}$ \and G.~Polenta$^{31}$ \and D.~Poletti$^{5}$ \and G.~Presta$^{1,2}$ \and S.~Realini$^{9,10}$ \and N.~Reyes$^{28}$ \and A.~Rocchi$^{18,19}$ \and J.~A.~Rubino-Martin$^{22}$ \and M.~Sandri$^{17}$ \and S.~Sartor$^{25}$ \and A.~Schillaci$^{15}$ \and G.~Signorelli$^{6}$ \and B.~Siri$^{11,12}$ \and M.~Soria$^{15}$ \and F.~Spinella$^{6}$ \and V.~Tapia$^{28}$ \and A.~Tartari$^{6}$ \and A.~Taylor$^{24}$ \and L.~Terenzi$^{17}$ \and M.~Tomasi$^{9,10}$ \and E.~Tommasi$^{31}$ \and C.~Tucker$^{4}$ \and D.~Vaccaro$^{6}$ \and D.~M.~Vigano$^{9,10}$ \and F.~Villa$^{17}$ \and G.~Virone$^{3}$ \and N.~Vittorio$^{18,19}$ \and A.~Volpe$^{31}$ \and B.~Watkins$^{24}$ \and A.~Zacchei$^{25}$ \and M.~Zannoni$^{8,23}$}

\institute{$^{1}$Univ. Roma Sapienza, 
$^{2}$INFN--ROMA1, 
$^{3}$IEEIT--CNR Torino, 
$^{4}$Cardiff Univ., 
$^{5}$SISSA Trieste,
$^{6}$INFN--Pisa,
$^{7}$INAF--IASF Milano,
$^{8}$Univ. Milano--Bicocca,
$^{9}$Univ Milano statale,
$^{10}$INFN--Milano,
$^{11}$Univ. Genova,
$^{12}$INFN--Genova,
$^{13}$IFAC--CNR Firenze,
$^{14}$Univ. Pisa,
$^{15}$Caltech,
$^{16}$Univ. Pennsylvania,
$^{17}$INAF Bologna,
$^{18}$Univ. Tor Vergata,
$^{19}$INFN--ROMA2,
$^{20}$Altran Italia Spa.,
$^{21}$JPL,
$^{22}$IAC Tenerife,
$^{23}$INFN--Milano Bicocca,
$^{24}$Oxford Univ.
$^{25}$INAF--OAT Trieste,
$^{26}$Univ. Padova
$^{27}$Manchester Univ.,
$^{28}$Univ. de Chile,
$^{29}$Univ. Ferrara,
$^{30}$INFN--Ferrara,
$^{31}$ASI
$^{32}$Univ. of Illinois
\\$^{*}$ \email{luca.lamagna@roma1.infn.it}}

\maketitle
\begin{abstract}

The Large Scale Polarization Explorer (LSPE) is a cosmology program for the measurement of large scale curl--like features (B--modes) in the polarization of the Cosmic Microwave Background. Its goal is to constrain the background of inflationary gravity waves traveling through the universe at the time of matter--radiation decoupling. The two instruments of LSPE are meant to synergically operate by covering a large portion of the northern microwave sky. LSPE/STRIP is a coherent array of receivers planned to be operated from the Teide Observatory in Tenerife, for the control and characterization of the low--frequency polarized signals of galactic origin; LSPE/SWIPE is a balloon--borne bolometric polarimeter based on 330 large throughput multi--moded detectors, designed to measure the CMB polarization at
\SI{150}{GHz} and to monitor the polarized emission by galactic dust above \si{200}{GHz}. The combined performance and the expected level of systematics mitigation will allow LSPE to constrain primordial B-modes down to a tensor/scalar ratio of $10^{-2}$.
We here report the status of the STRIP pre--commissioning phase and the progress in the characterization of the key subsystems of the SWIPE payload (namely the cryogenic polarization modulation unit and the multi--moded TES pixels) prior to receiver integration.


\end{abstract}

\section{Introduction}
Measurements of the Cosmic Microwave Background (CMB) polarization anisotropies offer strong probes of the standard cosmological model. Linear polarization of the CMB is known to emerge from Thomson scattering off electrons at the last scattering surface in the presence of local quadrupole anisotropies in the scattered radiation field. 
The polarization pattern observed in the sky is commonly decomposed in a curl--free “E--mode” and a divergence--free “B--mode”. E--modes alone emerge in the presence of the velocity fields around peaks in the baryon--photon fluid at last scattering, and thus exhibit a tight correlation with the temperature anisotropies of the CMB. As such, the TE angular cross--power spectrum and the EE power spectrum can be usefully combined with the TT power spectrum to break degeneracies among cosmological parameters, and to constrain specific parameters (e.g. the reionization optical depth) otherwise loosely constrained by temperature anisotropies. On the other hand, B--modes at small angular scales are observed as a result of E--to--B leakage due to gravitational lensing, and as such observed in correlation to the line--of--sight integrated lensing potential due to the presence of large scale structure across the light path to last scattering surface. 
In addition, a large (degree and above) scale contribution to the B--mode angular power spectrum is predicted to emerge from a stochastic background of tensor (metric) perturbations, i.e. gravitational waves, generated during the inflationary expansion history of the early universe. 
A rich variety of physical mechanisms can be devised to fit the inflationary paradigm to the needs of the standard cosmological scenario and to its current observational implant. The amplitude of the gravitational wave background, traced through its B--mode signal in the CMB polarization, can be effectively used to constrain such scenarios. The B--mode power yields a direct measurement of the so--called tensor--to--scalar ratio $r$, i.e. the power ratio between the tensor and scalar perturbations at the end of inflation.  A detection of large scale B--modes, if any at all, would thus represent the ultimate constraining probe for the physics of inflation. 

The best current upper limits on $r$, from the joint BICEP2 and Planck analysis, yields $r<0.07$ at $95\%$ C.L. \cite{joint}.
Observationally, the inflationary signal is expected to be at most a fraction of percent of the total polarized CMB power, invoking the need for large and ultra--sensitive focal planes. In addition, no region of the microwave sky can be safely considered free from foreground contamination down to the level required for primordial B--mode detection. 
Therefore, a large scale B--mode measurement demands very clever and careful strategies, both in instrument design and signal conditioning. While aiming at the mitigation of hardware systematics, it is also necessary to develop accurate modeling and treatment of the foregrounds through component separation techniques. Ideally, every spurious effect should be understood and traced across the analysis pipeline down to a fraction of the sensitivity level allowed by observations. 

The CMB community is currently devoting a huge effort towards large scale polarization measurements. A number of ongoing and upcoming projects, including the recently JAXA--approved LiteBIRD satellite \cite{LiteBIRDa, LiteBIRDb}, have been designed to shed light on the so--far elusive B--mode signal from inflationary gravitational waves.
\section{The Large Scale Polarization Explorer (LSPE)}
\begin{figure}[!htb]
   \centering
   \includegraphics[width=0.7\linewidth]{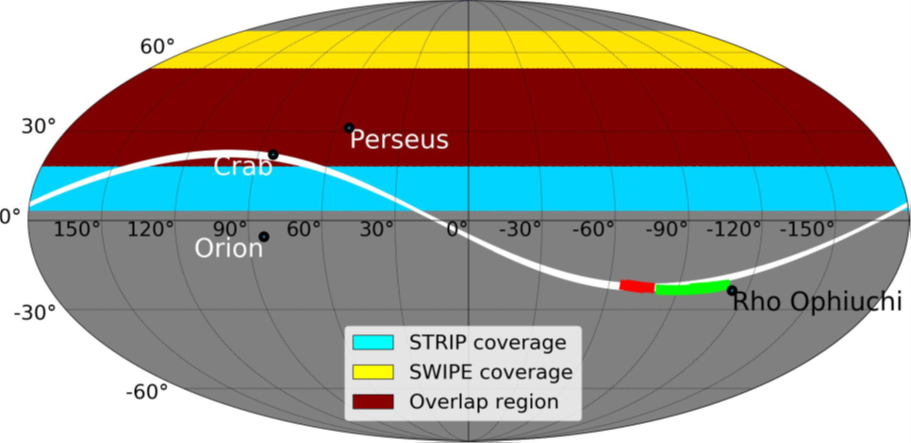}
   \caption{Sample of the combined sky coverage of the LSPE instruments (equatorial coordinates.)}\label{fig:lspecoverage}
\end{figure}

LSPE \cite{LSPE} is a program funded by the Italian Space Agency and INFN with the aims to:
\begin{itemize}
    \item perform a large--area survey of the microwave polarized sky in 5 spectral bands, with the scope to understand the galactic foregrounds at degree scales;
    \item test and validate hardware solutions, observing strategies, and data processing techniques for the observational challenge of the inflationary B--modes, in view of even more ambitious ground--based and space--borne programs devoted to CMB polarization in the next decade (including LiteBIRD);
    \item constrain $r$ at the level of $10^{-2}$;
    \item improve constraints on reionization and large scale CMB anomalies through combined measurements of temperature and polarization anisotropies at low multipoles.
\end{itemize}
LSPE will observe a 25\% fraction of the sky in the northern hemisphere by means of its two instruments, STRIP and SWIPE, complementary for frequency coverage and technology. 
The main characteristics of both instruments are summarized in Table~\ref{tab:features} and the combined sky coverage is shown in Fig.~\ref{fig:lspecoverage}.
\begin{table}[!htbp]
    \centering
    \fontsize{0.26cm}{0.32cm}\selectfont{
    \begin{tabular}{l |c c |c c c}
        \hline
        \hline
        Instrument & \multicolumn{2}{c|}{STRIP} & \multicolumn{3}{c}{SWIPE} \\
        \hline
        Site & \multicolumn{2}{c|}{Tenerife} & \multicolumn{3}{c}{balloon}\\
        Freq (GHz) & 43 & 90 & 145 & 210 & 240 \\
        Bandwidth & 16\%  & 8\%   &  30\%  & 20\%  & 10\% \\
        Angular resolution FWHM (arcmin) & 20 & 10  & \multicolumn{3}{c}{85}\\
        Detectors technology & \multicolumn{2}{c|}{HEMT}& \multicolumn{3}{c}{TES multimoded}\\
        Number of detectors $N_\text{det}$& 49 & 6 & 110 &108 & 108\\
        Detector NET  ($\mu \text{K}_\text{CMB}\sqrt{\text{s}}$)& 460 & 1247 &12.7 & 15.7 & 30.9\\
        Mission duration &  \multicolumn{2}{c|}{2 years} &  \multicolumn{3}{c}{15 days}\\
        Duty cycle  &  \multicolumn{2}{c|}{35\%}&  \multicolumn{3}{c}{90\%}\\
        Sky coverage $f_\text{sky}$& \multicolumn{2}{c|}{37\%}&  \multicolumn{3}{c}{38\%}\\
        Map sensitivity $\sigma_{Q,U}$ ($\mu \text{K}_\text{CMB}\cdot \text{arcmin}$)& 104 & 809 & 12&  14 & 29  \\
        Noise power spectrum $(\mathcal{N}^{E,B}_\ell)^{1/2}$ ($\mu \text{K}_\text{CMB}\cdot \text{arcmin}$)&171 &1330 &19&24&47\\
        \hline
        \end{tabular}
    }
    \caption{LSPE instrumental parameters.
    \label{tab:features}}
\end{table}
\section{STRIP}
STRIP (Survey TeneRife Italian Polarimeter) is a coherent polarimeter array that will observe the microwave sky from Tenerife in two frequency bands centered at 43 GHz (Q band, 49 receivers) and 95 GHz (W band, 6 receivers) through a dual--reflector crossed--Dragone telescope of 1.5 m projected aperture.
Formerly designed as a balloon--borne instrument to be flown along with SWIPE, STRIP has now been redesigned as a ground--based platform  that will start operations by Spring 2021 at the Teide Observatory in Tenerife. The STRIP main instrumental parameters for the two channels in the Q and W bands are summarized in Table 1. The Q--band goal sensitivity corresponds to an improvement of a factor 5 over the Planck--LFI 44 GHz sensitivity on the same pixel size at the end of the 30 months mission. The W--band is intended as an atmospheric emission and fluctuation monitor and will complement the Q--band measurements by allowing signal cleaning.
STRIP will scan the sky with a continuous azimuthal spin at 1 rpm, while maintaining telescope boresight at a fixed elevation angle (about 20 degrees from the zenith at Tenerife latitude). By combining the telescope spinning with the daily rotation of the Earth around its axis, STRIP will cover a region of the northern sky that overlaps at least with $80\%$ of the SWIPE survey.  
During daytime STRIP will manage the presence of sunlight by discarding data where the sun is at less than 10 degrees angular distance from the telescope line-of-sight. Our simulations show that this fraction corresponds to about $15\%$ of the data and it is included in our duty cycle computation.
When the sun is farther than 10 degrees from the boresight, its emission will be detected by the beam far sidelobes that are at the level of about \SI{-100}{dB}, enough to dilute this signal to negligible levels.

\begin{figure}[!htb]
   \begin{minipage}{\textwidth}
     \centering
     \includegraphics[width=\linewidth]{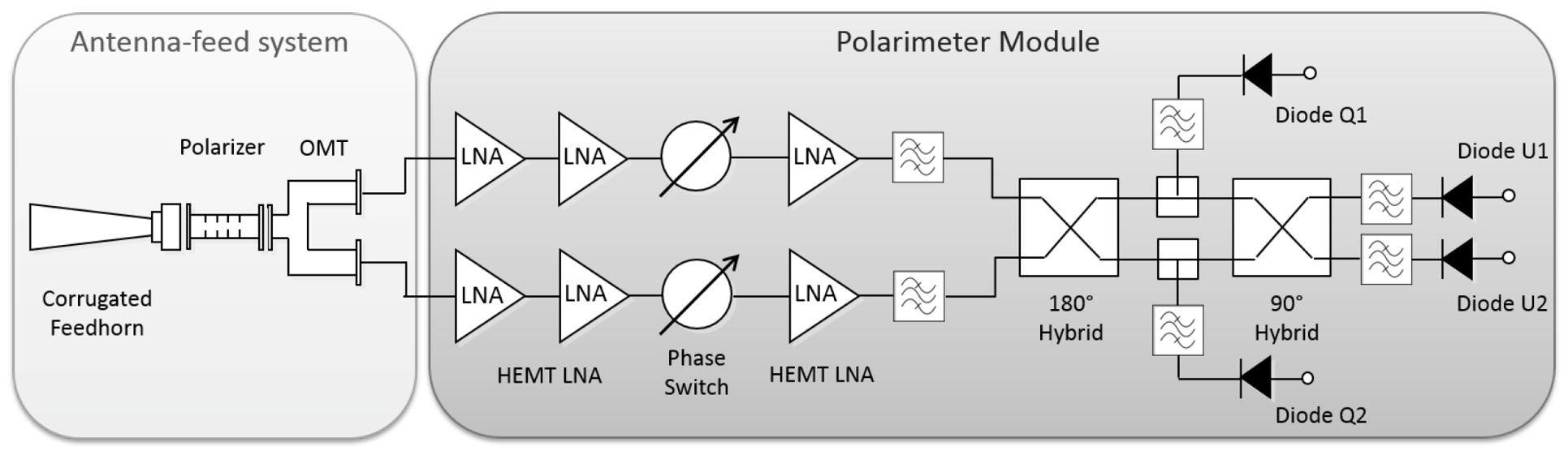}
   \end{minipage}\hfill 
   \caption{Schematic of the STRIP detection chain. The correlation unit receives the circular polarization coming from the orthomode transducer; each input is amplified and phase modulated by phase shifters. One phase switch shifts the phase between \SI{0}{\degree} and \SI{180}{\degree} at \SI{4}{kHz}, while the other works at \SI{100}{Hz}. The two signals are combined in a \SI{180}{\degree} hybrid coupler. Half of each output is bandpass filtered and rectified by a pair of GaAs detector diodes, while the other half passes through a \SI{90}{\degree} hybrid coupler. A second pair of bandpass filters and detector diodes measures the power coming out from the coupler.}\label{fig:stripschematics}
\end{figure}

The design of the STRIP polarimeters is based on InP cryogenic High Electron Mobility Transistor (HEMT) low noise amplifiers integrated in Monolithic Microwave Integrated Circuits (MMIC) and on high--performance waveguide components cooled to \SI{20}{K}. 
A schematic of the detection chain of STRIP is shown in Fig.~\ref{fig:stripschematics}. The benefit of this design is that the Q and U Stokes parameters are directly extracted from each pair of diodes at the output. Fig.~\ref{fig:stripschematics} shows  the principle of operation of its radiometers.

\begin{figure}[!htb]
   \begin{minipage}{0.5\textwidth}
     \centering
     \includegraphics[width=\linewidth]{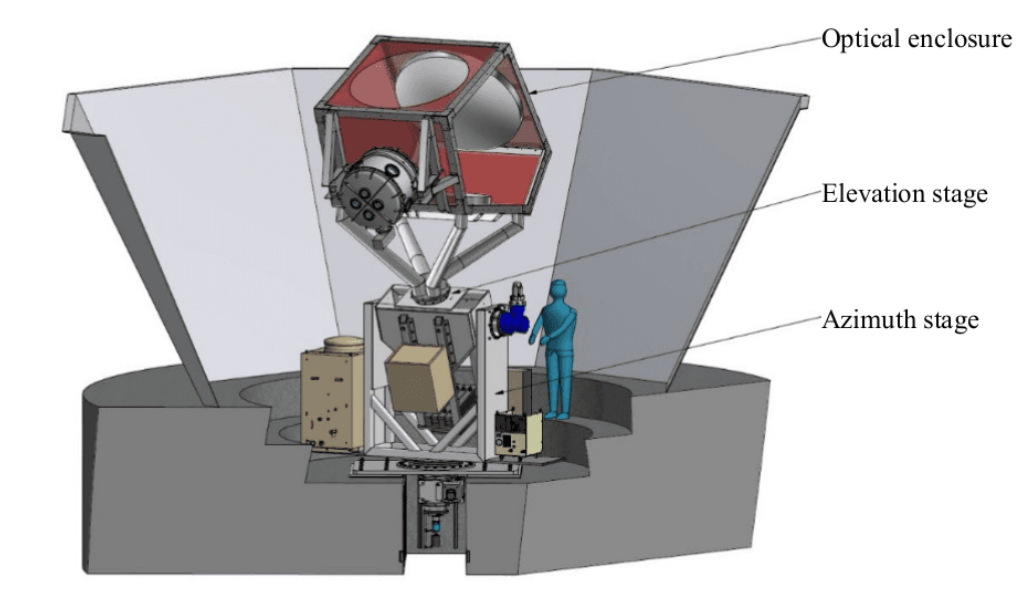}
   \end{minipage}\hfill
   \begin{minipage}{0.42\textwidth}
     \centering
     \includegraphics[width=\linewidth]{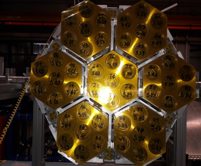}
   \end{minipage}\hfill 
   \caption{{\it Left:} STRIP telescope assembly. {\it Right:} Q--band corrugated feedhorns (incident photon view);}\label{fig:stripfoto}
\end{figure}

Fig.~\ref{fig:stripfoto} shows a sketch of the STRIP telescope and mount and a picture of the Q-band focal plane.

\section{SWIPE}
\begin{figure}[!htb]
   \begin{minipage}{0.48\textwidth}
     \centering
     \includegraphics[width=\linewidth,keepaspectratio]{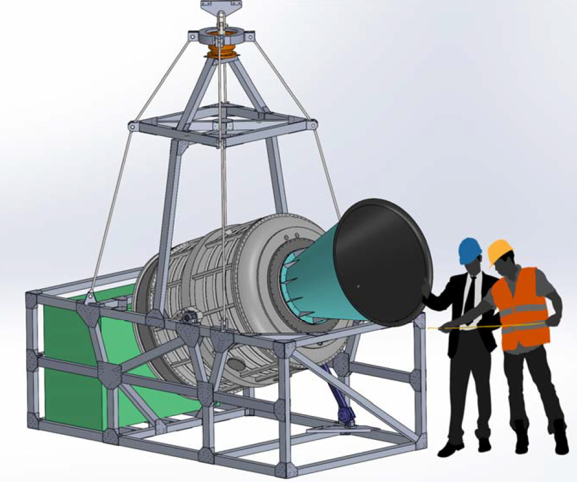}
   \end{minipage}\hfill
   \begin{minipage}{0.48\textwidth}
     \centering
     \includegraphics[width=\linewidth,keepaspectratio]{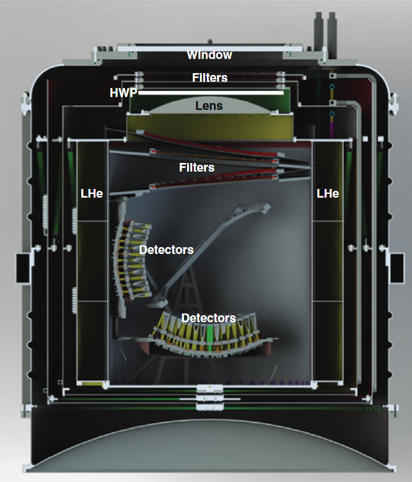}
   \end{minipage}\hfill 
   \caption{{\it Left:} sketch of the SWIPE gondola and receiver. {\it Right:}  schematic of the SWIPE receiver and optical system. }\label{fig:swipesketches}
\end{figure}

SWIPE (Short Wavelength Instrument of the Polarization Explorer) is a bolometric polarimeter based on a 50 cm refractive telescope. Its dual polarization focal plane hosts two arrays, one per polarization, which results in a total of 330 large throughput multi--moded TES bolometers operated at \SI{0.3}{K}, with passbands centered at 145, 210 and \SI{240}{GHz}.
SWIPE will survey the northern sky from a spinning long duration stratospheric balloon, which will be launched from high latitudes during the Arctic winter. Its main features are summarized in Table~\ref{tab:features}. 
Fig.~\ref{fig:swipesketches} shows a sketches of the SWIPE gondola, receiver and optical system. 
The key unique properties of SWIPE are the use of a large aperture half--wave plate (HWP), acting as a broadband polarization modulator at the telescope input (see e.g. \cite{Johnson2017},\cite{Columbro2019}), and the optimization of throughput with the multi--moded coupling of the single detectors to the system optics \cite{Legg}. 
The HWP rotation mechanism is based on a superconducting magnetic suspension system allowing continuous rotation at \SI{60}{rpm}, corresponding to \SI{4}{Hz} modulation of the polarized signal \cite{Columbro2019}. The stator of the modulation unit, hosting the superconducting magnets, is thermally attached to the pumped liquid He reservoir, ensuring an operating temperature below \SI{1.8}{K}. Magnetic friction is optimized to ensure that the rotor temperature does not rise above \SI{20}{K} under nominal operation.

\begin{figure}[!htb]
\begin{minipage}{\textwidth}
     \centering
     \includegraphics[width=\linewidth]{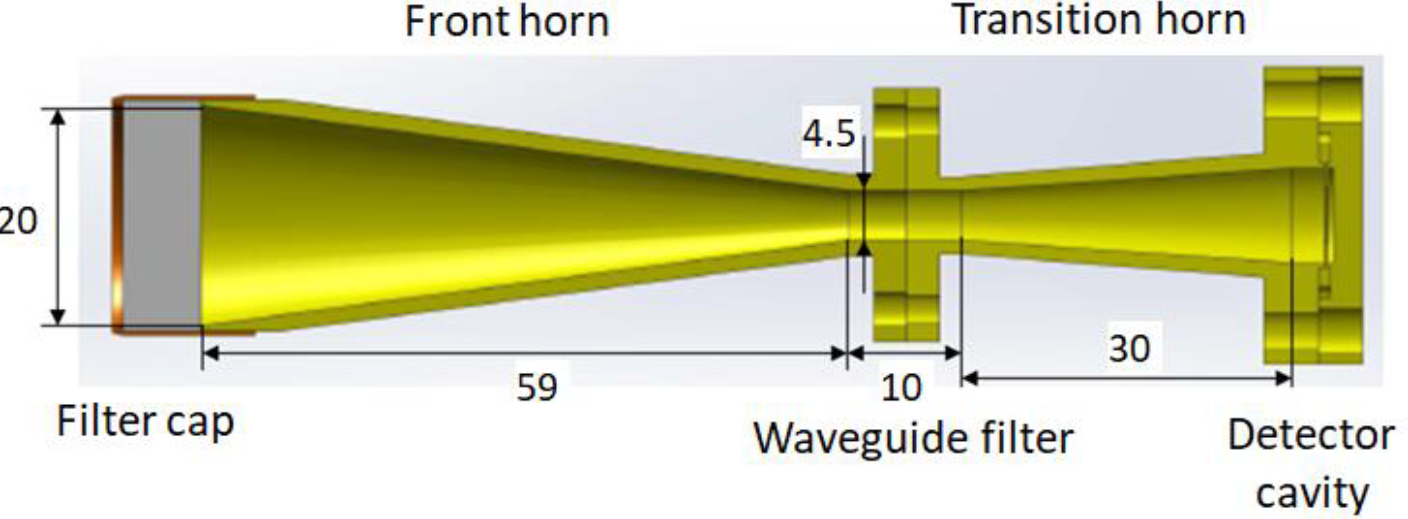}
   \end{minipage}\hfill   
   \begin{minipage}{0.24\textwidth}
     \centering
     \includegraphics[width=\linewidth]{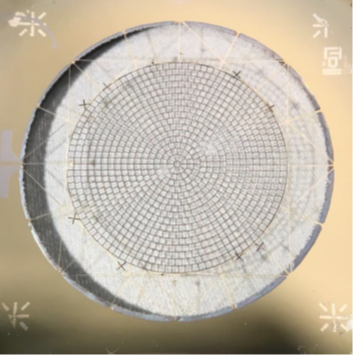}
   \end{minipage}\hfill
   \begin{minipage}{0.36\textwidth}
     \centering
     \includegraphics[width=\linewidth]{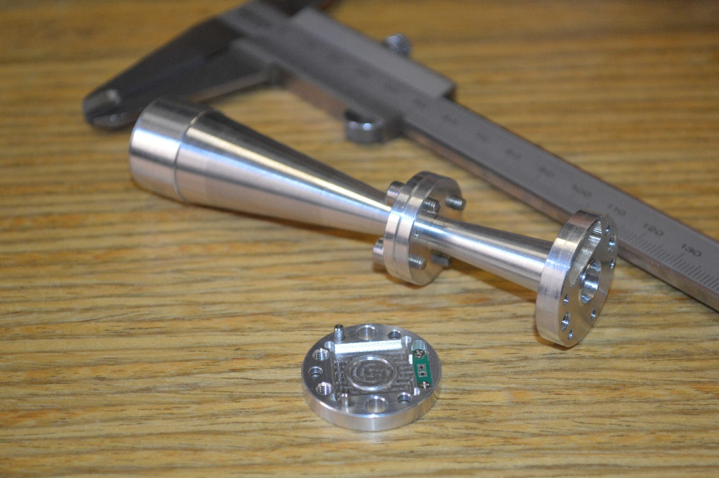}
   \end{minipage}\hfill 
   \begin{minipage}{0.34\textwidth}
     \centering
     \includegraphics[width=\linewidth]{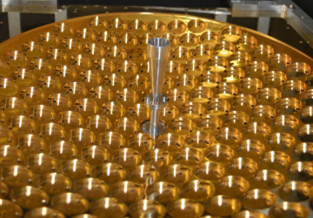}
   \end{minipage}\hfill
   \caption{{\it Top:}SWIPE multi-moded pixel assembly. The main components are highlighted and the quoted sizes are in mm.  {\it Bottom left:} SWIPE multi--moded TES bolometer absorber. {\it Bottom center:} SWIPE pixel assembly and bolometer cavity. {\it Bottom right:} sample pixel assembled on one of the two SWIPE focal planes.}\label{fig:swipepixels}
\end{figure}

The multi--moded detectors for SWIPE trade angular resolution for optical efficiency, by coupling few tens of modes per detector across the covered bands, so that the effective photon noise limited sensitivity equals that of a receiver with few thousands single--moded pixels in both polarizations \cite{Legg}. The broad and flat beam pattern of the pixels ensures a uniform illumination of the cold aperture stop of the telescope (with a -10dB edge taper at \SI{150}{GHz}). This is located just after the lens on the optical path from the sky to the focal planes (see Fig.~\ref{fig:swipesketches}). The larger pickup on the edges of the detector field of view does not hinder detector performance, since the telescope tube and the stop are cooled to less than \SI{2}{K}. In addition, highly absorptive coating is applied inside the tube in order to prevent in-band stray-light from leaking into the main beam.
The \SI{0.3}{K} pixel assembly (Fig.~\ref{fig:swipepixels}) is designed to couple each detector to the focal plane through a smooth--walled conical horn feeding a mode--filtering waveguide. After that, radiation propagates through a conical transition into a multi--moded cavity and it is detected by one Ti--Au TES sensor with a large (\SI{8}{mm} diameter) Bi--Au spider--web absorber \cite{Siri}. In SWIPE, polarization separation takes place at the level of the large polarizer splitting the beam to the two focal planes. Therefore, the polarization properties of the individual pixel assemblies do not affect the polarimeter performance.
The comparatively limited number of physical detectors, and of the bias/readout units needed for their operation, allows the receiver to comply with the strong power supply constraints due to the absence of the sun during the night flight, while still achieving a sensitivity scaling like the square root of the total number of coupled modes (see Table~\ref{tab:features}).

Implementation of multi--moded coupling implies obvious challenges in detector design and fabrication, due to the large size of the SiN spider--web absorbers and to the limited parameter space for performance optimization of the TES sensors \cite{Gualtieri, Siri}. The two SWIPE focal planes are cooled at a base temperature of \SI{0.3}{K}. Under the nominal loading conditions, optimal tradeoff between noise and saturation power requires a superconducting transition temperature $T_c$ of \SI{0.5}{K}.

The readout electronics is based on a Frequency Domain Multiplexing (FDM) scheme, with each dc-SQUID sensing 16 TESs biased in the 200 kHz-1.6 MHz frequency range, for a total of 24 dc-SQUIDs to handle the two focal planes \cite{vaccaro}. The cold part of the electronics is based on LC filter boards (including the bias resistor) next to the focal planes, at \SI{0.3}{K}, while the SQUID is at \SI{1.6}{K}, in a shielded box. The warm readout custom board is based on an Altera Cyclone V SoC, and it features mezzanine DAC and ADC boards. 

\section{Present status}
At present, most of the STRIP hardware has been developed and tested. Q--band and W--band sub--systems, including passive components (feedhorns, polarizers, OMT’s) and HEMT--based coherent receivers, have been successfully characterized at unit--level. Integration and system--level testing of the focal plane arrays inside the instrument cryostat is planned to be performed within the next months. 
Electronics and ancillary systems, such as the internal calibrator and the star tracker, are being finalized and verified as well. Verifications of the telescope parts are in progress. The preparation of the site and the construction of the sliding roof will be finalized between 2019 and 2020. STRIP is scheduled to be fully integrated at the observation site in Spring 2021.

The main subsystems of the SWIPE receiver are under construction and testing. In particular, we highlight below the work in progress on the main subsystems.
\begin{itemize}
    \item The clamp/release system for the magnetically levitating rotor has been successfully tested at cryogenic temperatures \cite{Clamp}. The other components of the HWP modulator have been built and individually tested, with the fully assembled system scheduled for cryogenic testing in the next months. 
    \item Tests of the detector prototypes for the SWIPE focal plane with a coherent source at 128 GHz prove a good consistency of the pixel response with the modeled multi--moded performance \cite{Columbro}. More extensive, broadband testing at the system--level is advised by the large size of the focal plane and by the frequency dependent positioning of the pixel units across the focal plane.
    \item The first functional tests (e.g., tone generation) on the prototype readout flight model have been performed \cite{tartari}.
\end{itemize}

The pre--integration campaign and subsystem compliance review for SWIPE are scheduled at the end of 2019. The first launch opportunity for SWIPE is late Arctic winter 2020, and all the related activities are currently on schedule to comply with that time window.




\end{document}